# A Taxonomy for Virtual and Augmented Reality in Education

**J. Motejlek**
Research Student
Department of Chemical and Process Engineering
University of Surrey, Guildford, Surrey, UK
E-mail: j.motejlek@surrey.ac.uk

**E. Alpay**
Professor / Director of Learning and Teaching
Department of Chemical and Process Engineering
University of Surrey, Guildford, Surrey, UK
E-mail: e.alpay@surrey.ac.uk

## 1    INTRODUCTION

Both virtual reality (VR) and augmented reality (AR) have undergone considerable development in recent years. Even though it seems that we are still in a primitive technological stage, it is already recognised that VR/AR can provide exciting opportunities to support teaching and learning [1]. There have been numerous attempts to use this technology in education contexts [2], in most cases showing success [3]. Example include military training applications [4], engineering applications through VR laboratories [5], and history [6] and astronomy [7] education. The possibilities to use VR/AR transcend to other contexts, such as interactive performances, theatre, galleries, discovery centres and so on [8]. The advantage of VR as an experimental and educational tool is the ability to place the participant inside any scene with high degree of immersion [9]. However, there are also examples where educational application has only been partially successful, such as the use of 3D anatomy models in medical education [10] or skill transfer in VR based microsurgery training [11]. Greater understanding is needed as to the features of such applications that are especially conducive to student learning. More fundamentally though, clarity is needed on the classification of the tools to accurately describe e.g. function and design.

In this paper, a taxonomy for VR/AR in education is presented that can help differentiate and categorise education experiences and provide indication as to why some applications of fail whereas others succeed. Examples will be presented to illustrate the taxonomy, including its use in developing and planning two current VR projects in our laboratory. The first project is a VR application for the training of Chemical Engineering students (and potentially industrial operators) on the use of a physical pilot plant facility. The second project involves the use of VR cinematography for enacting ethics scenarios (and thus ethical awareness and development) pertinent to engineering work situations.

## 2 CLASSIFICATION OF AR/VR IN EDUCATION

Key factors of the VR/AR taxonomy can be summarised as: (i) Purpose of the application, (ii) User experience, (iii) Technology of the delivery, (iv) Production technology, (v) Gamification type, (vi) User interaction and (vii) System interaction. A description of each of these factors is given below.

### 2.1 Classification by Purpose

The most important category in the taxonomy depends on the nature of the information being accessed and the intended purpose of this information. Specifically, purpose may involve:

#### A. Training

For training purposes the goal of the application is to convey information about how to use a specific real device (especially in case of AR) or its digitised equivalent (as in the case of model-based VR and cinematic VR). It is usually very specific in purpose, with the focus on training for equipment, machine or process operation rather than the understanding of the underlying principles of design. There are numerous examples of VR/AR training applications in education at the moment, extensively employed in medical training such as dentistry [12], laparoscopy [13] and ophthalmoscopy [14].

#### B. Teaching

For teaching purposes the goal is to prepare the student to retain and understand knowledge in a general situation. The student is being exposed to theory and underlying principles, and such knowledge is expected to be transferable to other situations and environments. Currently, the number of examples of successful teaching applications is relatively low, as often their development is challenging [15] with success relying on effective scaffolding [16] as well as effective integration of assessment and feedback. However, some examples of effective applications exist in areas of language teaching [17], general lab work [5] and agriculture [18].

#### C. Observing

For observing purposes, the primary goal is to show or convey information without the need for retaining or understanding it. In other words an *exhibition* purpose. Examples of such applications can be seen in the form of historical recreations of sights [6] or artefacts [19], the latter allowing for example shared analysis and research of objects between universities. The development of 3D scanners and video has played a key role for such observing purposes.

## 2.2 Classification by user experience and delivery technology

The two factors of *user experience* and *delivery technology* are closely related and are discussed together in this section. Currently, the taxonomy identifies three distinct user experiences:

### A. Virtual Reality (VR) Experience

VR experience completely isolates the user from the outside environment inside an immersive world [20]. The experience can therefore transport the user into both reality-simulated and hypothetical environments.

### B. Augmented Reality (AR) Experience

AR systems combine (overlay) virtual content (e.g. generated through a model, animation or video recording) with real-world imagery [21]. This occurs in real time as the user engages with the system, with aspects of the surroundings or other real-world objects registered in 3D [22]. In this way, AR can be used to enhance the real-word interaction and learning experience, helping to e.g. better elucidate principles and concepts.

### C. Display Experience

A virtual world can be presented on a standard 2D screen / display, as well as through 3D visualisation. Although this paper (and current technology development) focuses on the latter due to the immersive experience potential, a standard display experience does offer some advantages over 3D immersion, such as the relative clarity of text and general reading experience, and much less prone to causing user dizziness, headache or eyestrain [23].

The user experience may be delivered through two distinct hardware devices: the screen and a stereoscopic head mounted display (HMD). Specific features of these delivery devices are summarised below:

### A. HMD

HMD allows stereoscopic vision in VR [24] as well as in AR [20]. Stereoscopic vision arises when two views of the same scene with binocular disparity are presented to each eye. The effect depends on binocular fusion in order to yield perception of depth [25]. In both cases, the user is also hands-free, i.e. the user does not have to hold the device in their hands.

### B. Screen

The user uses a stationery (e.g. desktop computer) or hand-held (e.g. tablet) device [26], or may in fact be surrounded by the screen as in the case of the CAVE system [27].

Consideration of both the visual experience and the delivery technology creates the VR/AR technology matrix shown in Table 1; demonstrative examples of how current commercial VR/AR equipment are categorised within this matrix are also shown. The delivery and experience are of course closely related: the type of experience defined by the technology of delivery and to some extent its production methods (see below).

**Table 1**: User's experience vs. delivery technology

|  | **Screen** | **HMD** |
|---|---|---|
| **Display** | computer monitor | Google Glass |
| **VR** | simulators; panoramic videos | Vive; Oculus |
| **AR** | iPad | Meta; HoloLens |

Interestingly, the definition of AR as a 3D registered system [22] has a consequence for certain types of devices, such as Google Glass which, according to this taxonomy is actually a display experience, even though it resembles an AR experience. Since Google Glass does not show information related to what is in front of the user, nor is registered in 3D (Google Glass technology does not have the necessary sensors for that), it is basically a screen showing information similar to what the smart phone does but is strapped onto the user's head. Therefore, it should not be considered an AR device.

## 2.3 Classification by Production Technology

The production technology defines to an extent the type of delivery technology. In this taxonomy, it has been identified that it is possible to produce VR/AR experiences with 3D modelling, cinematography or combination of both.

### C. 3D Modelling

3D modelling and generated computer graphics is the most common approach to develop computer games and by extension *serious games* for education (see section 2.4 below). 3D models can be designed using tools, such as Blender or using photogrammetry using 3D scanners [19].

### D. Cinematography

In terms of cinematography, the footage is filmed with a specific field of view, most commonly 180 or 360 degrees, which then affects how much the user is surrounded by the image. The footage may also be filmed stereoscopically, i.e. to provide the illusion of 3D.

### E. Mixed

The two approaches can also be mixed. Indeed, embedding 3D objects within filmed footage is a common technique in the film industry. A blended approach with a delivery method that is screen based is not novel. However, the use of HMD for stereoscopic footage, that has also been enhanced with 3D models, is a new approach, and provides much design potential for educational tools. It should be noted that theoretically it is also possible to embed cinematography in a 3D model, but the authors know no such current applications. Such video embedment could involve for example the teacher or real process or equipment footage.

## 2.4 Gamification

Gamification describes game-inspired techniques to engage students within the learning / interaction process. The purpose of gamification is to increase student motivation for learning or skills development. In order to categorise the different types of gamification, intrinsic and extrinsic motivation concepts have been considered in this work, as well as the method of integration of gamification elements into the learning content.

Extrinsic motivation can be supported by rewards, and most gamification systems focus on this by using e.g. points, levels, leader boards, achievements or badges in order to motivate students to engage with learning content. The biggest disadvantage of this approach is that when the reward stops, the behaviour may also stop unless the student has found some other reason to continue. Reward based gamification is suitable for immediate and short term-change and has been observed to create a short term spike in user engagement [28].

Intrinsic motivation is the motivation which is driven by internal rewards and that do not depend on external controls, because they are perceived as inherently interesting and enjoyable by the student [29]. Research has shown that extrinsic rewards can undermine intrinsic motivation [30]. Nevertheless, some elements of extrinsic cue may help students monitor their level of progress through the learning activity, whilst not over-riding (or overwhelming) intrinsic drivers for learning.

It is possible to either embed game elements into the learning environment [31] or to integrate educational content into a game [32]. The latter are also referred to as *serious games*.

From this conceptual framework, two types of gamification can be derived:

### A. Reward based gamification

Adding elements such as leader boards, badges and achievements to the learning content in order to motivate students to progress through it. This on the whole may be seen as extrinsic motivators for the learning application.

### B. Serious games

Using game elements to increase students' internal motivation by adding educational content to the game.

According to [28] there are six elements inspired by game design, that can be used to increase intrinsic motivation within serious games: (i) mimicking play to facilitate the freedom to explore and fail within the boundaries of the game; (ii) the creation of stories for participants that are integrated with the real world; (iii) giving student's choices / options that then dictate the game plot; (iv) giving user information that connects concepts with real-world context; (v) encouraging participants to discover and learn from other interests in the real-world setting; and (vi) allowing participants to find connections to other interests and past knowledge within the game so as to deepen engagement and consolidate learning.

## 2.5 User Interaction

The design of the VR/AR application must also consider the methods of *user interaction* (e.g. information selection or exchange) with the user. This typically involve tracked controllers (e.g. gloves or sticks), or if these are not available or desirable, a simpler application control can be employed in form of gaze control [33], and are further defined below:

### A. Tracked controllers

Uses general-purpose controllers with buttons to interpret the user input. User points the controller in a direction and presses the button which causes the desired reaction from the system. Alternatively, the same effect can be achieved by tracking user's bare hands and interpreting gestures.

### B. Gaze control

User can see a cross hair in the middle of the viewport, and by moving his/her head can position the cross hair on the desired user interface element. Action is evoked either by pressing a button on the HMD or by waiting for certain amount of time (i.e. fixed gaze for a 1s or so).

### C. Special controllers

In terms of this taxonomy a special controller is a controller for input which cannot be replicated using a general-purpose controller and often employ precise simulated haptic feedback which helps students learn the required skill [13]. These controllers are common in medical training and can include for example virtual endoscopes [34], simulators of dental procedures [12] or ophthalmoscopes [14]. Such controllers can also be simpler, such as turning wheels for drivers [35].

## 2.6 System Interaction

The way the system communicates with the user is called *system interaction* in this taxonomy. It includes sophisticated subsystems embedded within the content, often based on research in artificial intelligence. It does not include basic menu and information that the application might include for the user to be able to operate its functionality. Two types of system interaction have been identified for inclusion in the taxonomy:

### A. Dialog systems

According to explanation-based constructivist theories of learning, learning is more effective and deeper when the learner must actively generate explanations than when merely presented with information [36]. This theory is being used by dialog systems, which ask the student to provide explanations of the educational context by means of menus or direct textual input. Effectiveness of learning is reported to be higher when the student is asked to answer questions via direct textual input [37]. Dialog systems are successfully used in non-VR/AR educational related applications [38] but they are not as easily employed in such form in VR/AR because it is harder to implement an effective method of input, especially when the experience is delivered via HMD [39].

### B. Intelligent agents

Intelligent agents are more sophisticated than dialog systems and interact with the user in a more complex way than just textual or audio information. The user can see their representation as an avatar which can move in the virtual space and operate objects in the virtual world [40], which adds life to the virtual world and improves immersivity of the VR application [41]. Intelligent agents can have the same effectiveness as human tutoring [42].

Interestingly, a lack of intelligent agents was identified as one of the problems of sustaining user immersion and interest in educational VR applications [43] and a number of authors planned to include such intelligent agents in future work (e.g. [44] and [6]). The intelligent agent can have at least three distinct functions that fall within the taxonomy:

a. Intelligent Agents for Training

   Shows how to perform tasks [40]. An example of this in tutoring is STEVE, which is an interactive autonomous system designed to teach students tasks and machinery operation related to naval engineering [40]. The agent recognises student's performance and can correct them in case they failed the task. The system also has ability to work in a team with more than one student [45].

b. Intelligent Agents for Teaching

   Explains abstract concepts [46]. Designing an explanation style of education is not trivial [15] and without preexisting scaffolding students may not be able progress in learning complex knowledge [16]. Software agents can have a significant influence on student motivation and it is important to ensure that agents facilitate, rather than dominate, the learning process [15]. Another factor to consider is importance of intentionality, orienting the learning activity around a problem-based teaching exercise which might promote a more intentional experience [15].

c. Intelligent Agents for Guiding

   This involves guiding students in a complicated environment that they are learning about so that they do not get lost (navigational guidance). The agent can also be an *attention guide* directing the student's gaze using pointing gestures [47]. In order to make the models and environments immersive the agents fulfil relevant tasks as if in the real world [6]. Such agents might not interact with the student and just be part of the simulation in order to increase immersivity.

### 2.7 Taxonomy overview

Based on the above review and discussions, an overview of the entire taxonomy is given in Table 2.

Table 2: Taxonomy Overview

| 1<br>Purpose | 2<br>Experience | 3<br>Production Technology | 4<br>Delivery Technology | 5<br>Gamification | 6<br>User Interaction | 7<br>System Interaction |
|---|---|---|---|---|---|---|
| 1.1<br>Training | 2.1<br>VR | 3.1<br>3D Modelling | 4.1<br>Screen | 5.1<br>Embedded game elements | 6.1<br>General purpose controller | 7.1<br>Dialog system |
| 1.2<br>Teaching | 2.2<br>AR | 3.2<br>Cinema-tography | 4.2<br>HMD | 5.2<br>Embedded educational content (serious games) | 6.2<br>Gaze control | 7.2<br>Intelligent Agents |
| 1.3<br>Observing | 2.3<br>Screen | 3.3<br>Mixed (rare) | | 5.3<br>None | 6.3<br>Special controller | 7.3<br>None |
| | | | | | 6.4<br>None | |

## 3 TAXONOMY APPLICATIONS

### 3.1 Chemical Process Pilot Plant Education Platform

A VR application of a Chemical process pilot plant has been developed in the VR lab of University of Surrey. Its purpose is for operation training of the actual plant within the same departments and providing a research platform for VR use in various educational settings. In terms of the taxonomy the platform can be used to develop applications for all three purposes, and so far, has been used for *training*. Specifically, the current application allows orienteering around the plant, helping students to understand the plant layout as well as recognize key items of equipment and instrumentation. The tool therefore enables safe and remote interaction with the plant. In terms of the developed taxonomy, the application can be classified as:

Purpose: Training, Experience: VR, Production Technology: 3D Modelling, Delivery Technology: HMD, Gamification: None, User Interaction: Tracked Controller, System Interaction: None

### 3.2 Stereoscopic Cinematography Storytelling

A second application has involved the set-up of a custom stereoscopic camera based on BlackMagic 4K cinema studio cameras for recording footage that can be viewed with HMD or on screen. The camera rig supports a viewing angle of 220º which eliminates all problems related to recording stereoscopic 360º videos. Since cinematography has higher potential for mobile device use due to lower hardware requirements, a gaze user interface was developed in order to provide interactivity to the video recordings.

The project has focussed on presenting students with a story related to chemical engineering, allowing them to make choices throughout the viewing, and eventually reaching an ethical dilemma near the conclusion of the story. The purpose of the learning interaction is therefore for ethical awareness within a professional / work context. In terms of the developed taxonomy, the application can be classified as:

Purpose: Observing, Experience: VR, Production Technology: Cinematography, Delivery Technology: HMD (both desktop and mobile grade), Gamification: None, User Interaction: Gaze, System Interaction: None.

## 4 DISCUSSION

For the applications described in section 3 above, future work will allow evaluation of their effectiveness with respect to the specific taxonomical features. Specific areas of development (and research evaluation) also arise through consideration of the taxonomy, including: (i) the use of mixed cinematography technology; (ii) the combined use of HMD and display delivery especially in group work situations (e.g. a student using the HMD and other group members viewing the student experience to facilitate additional discussion and reflection); and (iii) aspects of the application that would benefit from gamification and system interaction. Moreover, the taxonomy provides indication as to how the base applications can be evolved for other teaching / training / observing scenarios. For example, a greater teaching rather than training purpose may be created through the inclusion of relevant system interaction components.

In a related manner, it is also important to understand the critical definition of *purpose* in VR/AR application design. Imagine the following example: students are presented with a chemical engineering rig in VR, containing various instruments (filters, heaters, reactor, pumps, etc…), and are asked to locate these instruments. It might be that the student is learning what each instrument looks like, which would suggest the application purpose is *teaching*. It can be argued that the student is actually learning the positions of those instruments, which is specific for the given rig, and its purpose could therefore be *training* for subsequent plant operation. Whilst in real-word situations, both teaching and training elements may occur simultaneously (complemented by tutor / demonstrator input, reading material or lectures / tutorials), care is needed in VR/AR design to ensure the effective attainment of purpose through, e.g., appropriate production and delivery technologies, gamification and user and system interaction. This could arguably lead to an $8^{th}$ category in the taxonomy, stand-alone application vs. complementary resource. However, in Higher Education contexts, it is envisaged that most learning tools (especially in engineering) are not disparate to other teaching and learning experiences.

## 5 CONCLUSION

A seven-factor taxonomy for VR/AR in education has been presented. This has been constructed through consideration of current applications and literature, as well as consideration of aspects of application purpose, design, interaction and engagement. The taxonomy provides a framework for categorising and verbalising educational applications in VR/AR, as well as for identifying areas for specific (and novel) development and research evaluation.